\documentclass{natureprintstyle}
\usepackage{graphicx}
\usepackage{amssymb}

\date{October 20, 2016} 
\newcommand\farcm{\mbox{$^\prime$}}%

\title{Ultraluminous X-ray bursts in two ultracompact companions to nearby
elliptical galaxies}

\author{Jimmy A. Irwin$^{1}$, W. Peter Maksym$^{2}$, Gregory R. Sivakoff$^{3}$, Aaron J. Romanowsky$^{4,5}$, Dacheng Lin$^{6}$, Tyler Speegle$^{1}$, Ian Prado$^{1}$, David Mildebrath$^{1}$,
Jay Strader$^{7}$, Jifeng Liu$^{8,9}$, Jon M. Miller$^{10}$
}

\begin{document}

\maketitle

\begin{affiliations}
\item Department of Physics and Astronomy, University of Alabama, Box
  870324, Tuscaloosa, Alabama, 35487, USA.
\item Harvard-Smithsonian Center for Astrophysics,
60 Garden St., Cambridge, MA 02138, USA.
\item Department of Physics, University of Alberta, CCIS 4-181, Edmonton, AB T6G 2E1, Canada
\item Department of Physics \& Astronomy, San Jos\'e State University, One Washington Square, San Jose, CA 95192, USA.
\item University of California Observatories, 1156 High Street, Santa Cruz, CA 95064, USA
\item Space Science Center, University of New Hampshire, Durham, NH 03824, USA.
\item Department of Physics and Astronomy, Michigan State University, East Lansing, MI 48824, USA.
\item Key Laboratory of Optical Astronomy, National Astronomical Observatories, Chinese Academy of Sciences, 20A Datun Rd, Chaoyang District, Beijing, China 100012
\item College of Astronomy and Space Sciences, University of Chinese Academy of Sciences, 19A Yuquan Road, Beijing 100049, China
\item Department of Astronomy, The University of Michigan, 1085 South University Avenue, Ann
Arbor, Michigan, 48103, USA.

\end{affiliations}

\textbf{An X-ray flaring source was found near the galaxy NGC 4697$^{1}$.
Two flares were seen, separated by four years. The flux increased by a factor of
90 on a timescale of about one minute. Both flares were very brief. There is no
optical counterpart at the position of the flares$^{1}$, but if the source was at
the distance of NGC 4697, the luminosities were $\mathbf{>10^{39}}$ erg s$^{-1}$.
Here we report the results of a search of archival X-ray data for 70 nearby
galaxies looking for similar such flares. We found two flaring sources in
globular clusters or ultra-compact dwarf companions of parent elliptical galaxies.
One source flared once to a peak luminosity of
$\mathbf{9 \times 10^{40}}$ erg s$^{-1}$, while the other flared five times to
$\mathbf{10^{40}}$ erg s$^{-1}$. All of the flare rise times were $<$1 minute,
and they then decayed over about an hour. When not flaring, the sources appear
to be normal accreting neutron star or black hole X-ray binaries, but they are
located in old stellar populations, unlike the magnetars, anomalous X-ray pulsars
or soft gamma repeaters that have repetitive flares of similar luminosities.}

One source (hereafter Source 1) is located at RA=12:42:51.4, Dec=+02:38:35 (J2000)
near the Virgo elliptical galaxy NGC~4636 ($d= 14.3$ Mpc)$^{2,3}$. The cumulative
X-ray photon arrival time plot and a crude background-subtracted light curve for
this source derived from a $\sim$76,000 second {\it Chandra} observation taken on
2003 February 14 are shown in Figure~1. Prior to and after the flare, the X-ray
count rate of the source was $2.1 \pm 0.2 \times 10^{-3}$ counts s$^{-1}$
corresponding to a 0.3--10 keV luminosity of
$7.9 \pm 0.8 \times 10^{38}$ erg s$^{-1}$ for a power law spectral model with a
best-fit photon index of $\Gamma = 1.6 \pm 0.3$ (see {\it Methods}) if the source
is at the distance of NGC~4636. About 12,000 seconds into the observation, the
source flared dramatically, with six photons detected in a 22 second span, leading
to a conservative peak flare count rate of $0.3^{+0.2}_{-0.1}$ counts s$^{-1}$, a
factor of 70--200 increase in emission over its persistent (non-flare) state.
Assuming the same spectral model as in the persistent state, the flare peaks at
$9^{+6}_{-4} \times 10^{40}$ erg s$^{-1}$. Following the initial 22 second burst,
the source emitted at a less intense but still elevated rate for the next
1,400 seconds. In total, 25 photons were emitted during the flare, for an average
X-ray luminosity of $7 \pm 2 \times 10^{39}$ erg s$^{-1}$, and a total flare
energy of $9 \pm 2 \times 10^{42}$ erg. We assess the probability of this burst
being due to a random Poisson fluctuation of the persistent count rate is
$\sim$$6\times 10^{-6}$ (see {\it Methods}). While the photon statistics during
the 25-photon burst were limited, there was no evidence that the spectrum of the
source differed during the flare. There are no apparent flares in the combined
370,000 seconds of the other {\it Chandra} and {\it XMM Newton} observations of
NGC 4636, both before or after 2003 February 14 (see Extended Data Table 1).

A previous study$^{4}$ associated Source 1 spatially with a purported globular
cluster of NGC~4636 identified through Washington $C$ and Kron-Cousins $R$
system CTIO Blanco Telescope imaging$^{5}$. While faint ($R$=23.02), the
optical source identified as CTIO ID 6444 had a $C-R = 1.94$ color consistent
with a globular cluster of near-solar metallicity$^{5}$. Follow-up spectroscopic
observations$^{6}$ of a sub-sample of the globular cluster candidates in the
vicinity of the globular cluster hosting Source 1 found that 52/54 (96\%) of the
objects with a $C-R$ color and magnitude similar to CTIO ID 6444 were confirmed
to be globular clusters of NGC~4636, with the two remaining objects identified
as foreground Galactic stars. The hard X-ray spectrum of Source 1 during its
persistent phase (see {\it Methods}) argues against it being a late-type Galactic
dwarf star, for which the X-ray emission tends to be quite soft. Galactic RS CVn
stars exhibit hard X-ray emission in quiescence and are known to undergo X-ray
flares, but RS CVn have much higher optical--to--X-ray flux ratios$^{7}$ compared
to Source 1. Thus, it is highly likely that the optical counterpart of
Source 1 is a globular cluster within NGC~4636. Based on its absolute $R$
magnitude and $M/L= 4.1$, we estimate the globular cluster to have a mass of
$3 \times 10^{5}$ M$_{\odot}$ (see {\it Methods}).

A second X-ray source located near the elliptical galaxy NGC~5128 showed similar
flaring behavior. In the 2007 March 30  {\it Chandra} observation of NGC~5128, a
source at RA=13:25:52.7, Dec=$-$43:05:46 (J2000; hereafter Source 2) began the
observation emitting at a count rate of
$9.5 \pm 1.5 \times 10^{-4}$ counts s$^{-1}$ corresponding to a 0.3--10 keV
luminosity of $4.4 \pm 0.7 \times 10^{37}$ erg s$^{-1}$ using the best-fit
$\Gamma=1.0 \pm 0.2$ power law photon index and a distance$^{8}$ of 3.8 Mpc for
NCG~5128. Midway through the observation, the source flared dramatically, with
10 photons detected in a 51 second time span corresponding to a conservative
peak luminosity estimate of $9^{+4}_{-3} \times 10^{39}$ erg s$^{-1}$, after
which the flare subsided. Following the flare, Source 2 returned to its
pre-flare luminosity for the remainder of the observation.

Inspection of other archival {\it Chandra} and {\it XMM-Newton} data (see
Extended Data Table 1) revealed four more flares of Source 2. Three were observed
with {\it Chandra} on 2007 April 17, 2007 May 30, and 2009 January 4,
respectively, while the fourth flare was observed with {\it XMM-Newton} on 2014
February 9. In each instance, during the initial fast ($<$30 seconds) rise of the
flare the count rate increased by a factor of 200--300 over the persistent count
rate to $\sim$$10^{40}$ erg s$^{-1}$, after which the flare subsided. The total
flare energy of each of the five flares was $\sim$$10^{42}$ erg. The light curves
for the four {\it Chandra} flares look remarkably similar, as illustrated in
Figure~2. We combined these four light curves into a combined background-subtracted
light curve (see {\it Methods} for details) in Figure~2. Following the fast rise of
the source by a factor $\sim$200 to a peak luminosity approaching
$10^{40}$ erg s$^{-1}$, the source remained in a roughly steady ultraluminous state
for $\sim$200 seconds before decaying over a time span of $\sim$4,000 seconds
(Figure~2). Fitting a power law to the combined spectra of the four {\it Chandra}
flares yielded a best-fit photon index of $\Gamma = 1.2 \pm 0.3$. Thus, much like
Source 1, the spectrum of Source 2 did not change appreciably during the flare.

Source 2 has been previously identified with the object HGHH-C21 (also called
GC 0320) within NGC~5128$^{9-11}$. With a spectroscopically-determined recessional
velocity$^{10}$ (460 km s$^{-1}$) within 110 km s$^{-1}$ of NGC~5128, the source is
clearly at the distance of NGC~5128. This implies a projected half-light radius$^{12}$
of 7 pc. With a velocity dispersion of 20 km s$^{-1}$ and an inferred stellar
mass$^{12}$ of $3.1 \times 10^6$ M$_{\odot}$, the optical counterpart is either a
massive globular cluster or, given its unusual elongated shape, more likely an
ultracompact dwarf companion galaxy of NGC~5128.

It is unlikely that the flaring and the steady emission in both sources can be
attributed to two unrelated sources in the same host. Since our flare search technique
would have found these flares had they been detected by their flare photons alone, we
can calculate the probability that these globular clusters would have also hosted
steady X-ray emission more luminous than the persistent emission in each globular
cluster (see {\it Methods}). The globular cluster in Source 1 has a $<$0.3\%
probability of having an X-ray source with a luminosity more than
$8 \times 10^{38}$ erg s$^{-1}$, while the globular cluster/ultracompact dwarf in
Source 2 has a $<$9\% probability of having an X-ray source with a luminosity more than
$4 \times 10^{37}$ erg s$^{-1}$. Multiplying these probabilities leads to only a
$<$0.02\% chance that both flares are unrelated to the steady emission.  Of course, in
the unlikely event that the flares are distinct sources from the persistent sources, the
flaring sources must be flaring by more than two orders of magnitude over whatever
their true non-flare luminosities are.

Summing up all the available archival {\it Chandra} and {\it XMM-Newton} data (but
omitting the {\it Chandra} HRC and transmission grating exposures, which are not
sensitive enough to detect a flare of similar intensity seen in the ACIS observations)
allows us to constrain the duty cycle and recurrence rate of the flares. Source 2 flared
five times for a total combined flare time of $\sim$20,000 seconds in a total observation
time of $7.9 \times 10^5$ seconds, yielding one flare every $\sim$1.8 days and a duty
cycle of $\sim$2.5\%. Source 1 flared once for 1,400 seconds in a total observation time
of 370,000 seconds. This single flare implies a recurrence timescale of $>$4 days and
duty cycle of $<$0.4\%. 

In terms of energetics, variability, and survivability, only short and intermediate
duration soft gamma repeaters (SGRs)$^{13}$ and their cousins the anomalous X-ray pulsars
(AXPs)$^{14}$ are comparable to the sources discussed here. However, SGRs/AXPs are
believed to be very young and highly magnetized neutron stars, which would not likely be
found in an old stellar population such as a globular cluster or red ultracompact dwarf
galaxy. Our sources are also unlike SGRs/AXPs in that SGR/AXP flares of this magnitude
only last a few to a few tens of seconds$^{15, 16}$ without an hour long decay as seen in
our sources. Our sources are also unlikely to be Type II X-ray bursts of neutron stars,
believed to result from rapid spasmodic accretion onto the neutron star. In addition to
having flare--to--pre-flare luminosity ratios of only 10--20, the only Type II burst to
reach $10^{40}$ erg s$^{-1}$ (GRO 1744-28 -- the Bursting Pulsar) exhibits several
sub-minute flares per day when flaring, with much lower total flare energies per burst than
our sources and different timing properties from our sources$^{17}$. Furthermore, the
quiescent X-ray luminosity of the Bursting Pulsar is 4--5 orders of magnitude fainter than
the long-term luminosities of our sources. Qualitatively, the fast rise--slower decay of
Source 2 (Figure~2) resembles that of Type I bursts from Galactic neutron stars, which
typically peak near the Eddington limit of a neutron star. However, the peak luminosities
from Sources 1 and 2 are 1--2 orders of magnitude greater than the Type I limit for even
helium accretion, and last more than an order of magnitude longer. Rare superbursts from
Galactic neutron stars have been known to last for an hour$^{18, 19}$ but also have peak
luminosities well below those of our sources. Other X-ray flares of unknown source published
in the literature$^{20-22}$ appear to be one-time transient events, indicating that they were
(most likely) cataclysmic events with no post-flare emission, unlike our sources.

We have investigated the light curves of several thousand X-ray point sources within
70 {\it Chandra} observations of nearby galaxies and found only these two examples. It would
appear that the Milky Way has no analogs to our sources. Given the number of X-ray sources in
the Milky Way brighter than $10^{37}$ erg s$^{-1}$ ($\sim$40)$^{23}$, the lack of X-ray
binaries more luminous than $10^{38}$ ergs s$^{-1}$ in Galactic globular clusters, and the
rarity of burst sources in the extragalactic sample, this is not surprising. The nature of
these sources remains uncertain. The increased emission might result from a narrow cone of
beamed emission crosses our line of sight every few days. It is unclear, however, how a pulsed
beam would lead to the distinctly asymmetric fast rise--slower decay profile. Alternatively,
the flare might represent a period of rapid highly super-Eddington accretion onto a neutron
star or stellar-mass black hole, perhaps during the periastron passage of a donor companion
star in an eccentric orbit around a compact object. Such a explanation has been suggested to
explain observed (albeit neutron star Eddington-limited) flares in galaxies$^{1, 24}$.
Finally, the high X-ray luminosity during the peak of the flare might represent accretion onto
an intermediate-mass black hole. If the flares are Eddington-limited, black hole masses of
800 M$_{\odot}$ and 80 M$_{\odot}$ are implied for Sources 1 and 2, respectively, assuming a
bolometric correction of 1.1 appropriate for a 2 keV disk blackbody temperature spectral model.
The fast rise times constrain the maximum mass of a putative black hole, since the rise time
cannot be shorter than the light travel time across the innermost stable circular orbit of the
black hole. For both sources, the fastest rise happened over a 22 second period, implying an
upper limit on the mass of a maximally-rotating black hole of $2 \times 10^6$ M$_{\odot}$.
A black hole in this mass range is a particularly intriguing explanation for Source 2 if indeed
its host is the stripped core of a dwarf galaxy.

\noindent \textbf{References}

\noindent [1.] Sivakoff, G. R., Sarazin, C. L. \& Jord\'an, A. Luminous X-Ray
Flares from Low-Mass X-Ray Binary Candidates in the Early-Type Galaxy NGC 4697.
Astrophys. J. 624, L17--L20 (2005).

\noindent [2.] Tonry, J. L. et al. The SBF Survey of Galaxy Distances. IV. SBF
Magnitudes, Colors, and Distances. Astrophys. J. 546, 681--693 (2001).

\noindent [3.] Mei, S. et al. The ACS Virgo Cluster Survey. XIII. SBF Distance
Catalog and the Three-dimensional Structure of the Virgo Cluster. Astrophys. J.
655, 144--162 (2007).

\noindent [4.] Posson-Brown, J., Raychaudhury, S., Forman, W., Donnelly,
R. H., \& Jones, C. Chandra Observations of the X-Ray Point Source Population
in NGC 4636. Astrophys. J. 695, 1094--1110 (2009).

\noindent [5.] Dirsch, B., Schuberth, Y. \& Richtler, T. A wide-field
photometric study of the globular cluster system of NGC 4636. Astron.
Astrophys. 433, 43--56 (2005).

\noindent [6.] Schuberth, Y. et al. Dynamics of the NGC 4636 globular cluster
system. An extremely dark matter dominated galaxy? Astron. Astrophys. 459,
391--406 (2006).

\noindent [7.] Pandey, J. C. \& Singh, K. P. A study of X-ray flares - II. RS
CVn-type binaries. Mon. Not. R. Astron. Soc. 419, 1219--1237 (2012).

\noindent [8.] Harris, G. L. H., Rejkuba, M. \& Harris, W. E. The Distance to
NGC 5128 (Centaurus A). Publ. Astron. Soc. Australia 27, 457--462 (2010).

\noindent [9.] Harris, G. L. H., Geisler, D., Harris, H. C., \& Hesser, J. E.
Metal abundances from Washington photometry of globular clusters in NGC 5128.
Astron. J. 104, 613--626 (1992).

\noindent [10.] Woodley, K. A. et al. The Kinematics and Dynamics of the
Globular Clusters and Planetary Nebulae of NGC 5128. Astron. J. 134, 494--510
(2007).

\noindent [11.] Woodley, K. A. et al. Globular Clusters and X-Ray Point
Sources in Centaurus A (NGC 5128). Astrophys. J. 682, 199--211 (2008).

\noindent [12.] Mieske, S. et al. On central black holes in ultra-compact
dwarf galaxies. Astron. Astrophys. 558, A14 (2013).

\noindent [13.] Kouveliotou, C. et al. An X-ray pulsar with a superstrong magnetic
field in the soft $\gamma$-ray repeater SGR1806-20. Nature 393, 235--237 (1998).

\noindent [14.] Mereghetti, S. \& Stella, L. The very low mass X-ray binary
pulsars: A new class of sources? Astrophys. J. 442, L17--L20 (1995).

\noindent [15.] Olive, J-. F. et al. Time-resolved X-Ray Spectral Modeling of
an Intermediate Burst from SGR 1900+14 Observed by HETE-2 FREGATE and
WXM. Astrophys. J. 616, 1148--1158 (2004).

\noindent [16.] Kozlova, A. V. R et al. The first observation of an intermediate
flare from SGR 1935+2154. Mon. Not. R. Astron. Soc. 460, 2008 (2016).

\noindent [17.] Younes, G. et al. Simultaneous NuSTAR/Chandra Observations of
the Bursting Pulsar GRO J1744-28 during Its Third Reactivation. Astrophys. J.
804, 43 (2015).

\noindent [18.] Cornlisse, R., Heise, J., Kuulkers, E., Verbunt, F., \& in't Zand,
J. J. M. The longest thermonuclear X-ray burst ever observed?. A BeppoSAXWide
Field Camera observation of 4U 1735-44. Astron. Astrophys. 357, L21--L24 (2000).

\noindent [19.] Strohmayer, T. E. \& Beown, E. F. A Remarkable 3 Hour Thermonuclear
Burst from 4U 1820-30. Astrophys. J. 556, 1045--1059 (2002).

\noindent [20.] Jonker, P. G. et al. Discovery of a New Kind of Explosive X-Ray
Transient near M86. Astrophys. J. 779, 14 (2013).

\noindent [21.] Luo, B., Brandt, W. N. \& Bauer, F. Discovery of a fast X-ray
transient in the Chandra Deep Field-South survey. The AstronomerÕs Telegram 6541,
1 (2014).

\noindent [22.] Glennie, A., Jonker, P. G., Fender, R. P., Nagayama, T., \&
Pretorius, M. L. Two fast X-ray transients in archival Chandra data. Mon. Not.
R. Astron. Soc. 450, 3765--3770 (2015).

\noindent [23.] Grimm, H.-J.., Gilfanov, M. \& Sunyaev, R. The Milky Way in X-rays
for an outside observer. Log(N)-Log(S) and luminosity function of X-ray binaries
from RXTE/ASM data. Astron. Astrophys. 391, 923--944 (2002).

\noindent [24.] Maccarone, T. J. An explanation for long flares from extragalactic
globular cluster X-ray sources. Mon. Not. R. Astron. Soc. 364, 971--976 (2005).

\noindent \textbf{Acknowledgments:}
We thank Tom Richtler for helpful discussions. J.A.I. was supported from
{\it Chandra} grant AR6-17010X and NASA ADAP grant NNX10AE15G. G.R.S. acknowledges
support of an NSERC Discovery Grant. A.J.R was supported by the National Science
Foundation grant AST-1515084. J.S. acknowledges support from NSF grants AST-1308124
and AST-1514763 and the Packard Foundation.

\noindent \textbf{Author Contributions:}
J.A.I. led the {\it Chandra} data reduction and analysis, with contributions from
W.P.M for the {\it XMM-Newton} data reduction and analysis. T.S., I.P., and D.M.
conducted the {\it Chandra} galaxy survey that yielded the two flare sources, with
oversight from J.A.I. G.R.S., A.J.R., D.L., J.S., J.L., and J.M.M. contributed to
the discussion and interpretation.

\noindent \textbf{Competing Interests:}
The authors declare that they have no competing financial interests.

\noindent \textbf{Correspondence:}
Correspondence and requests for materials should be addressed to
J.A.I. (jairwin@ua.edu).

\clearpage

\begin{figure*}
\begin{center}
\includegraphics[width=3.2in]{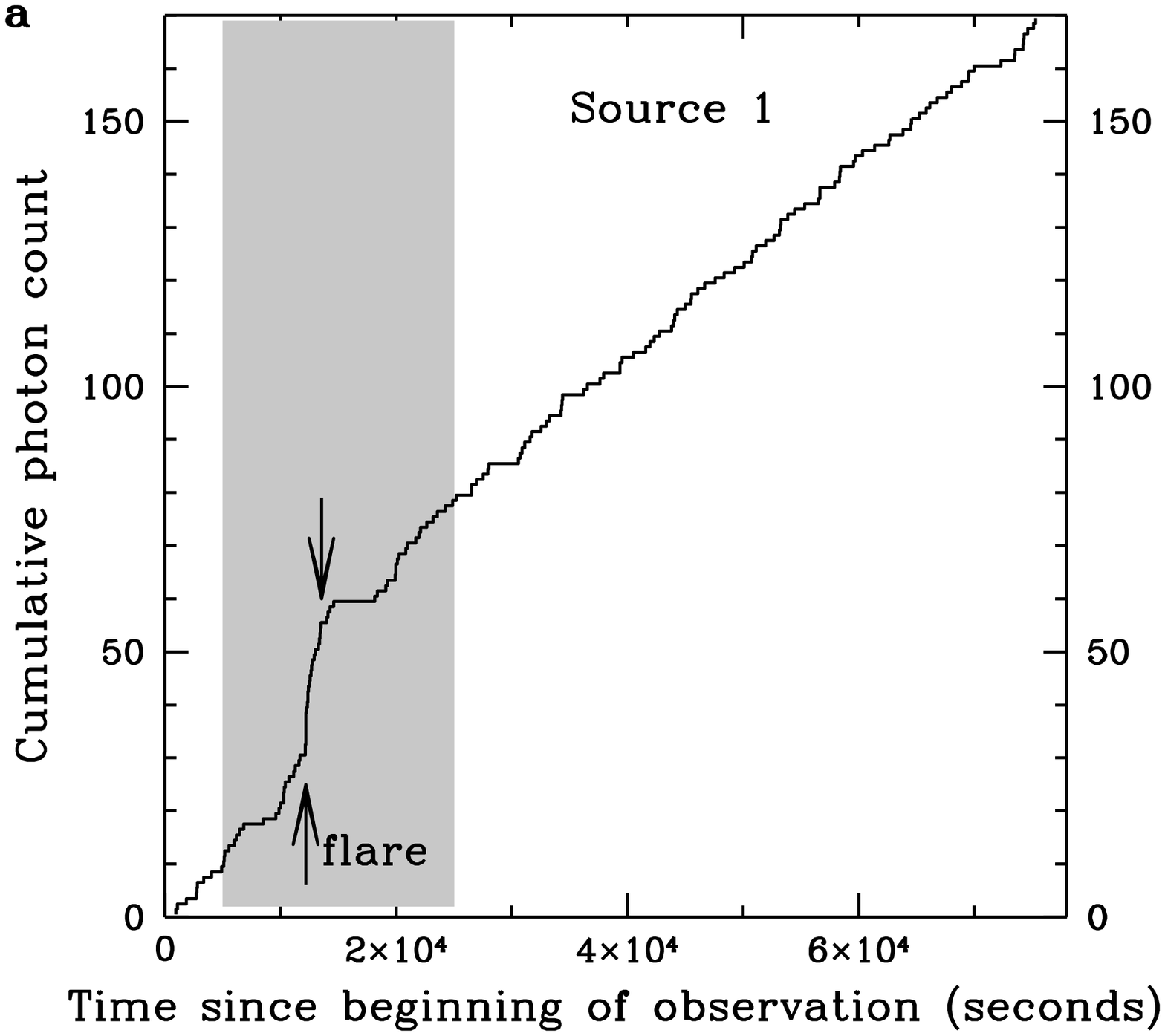}
\includegraphics[width=3.2in]{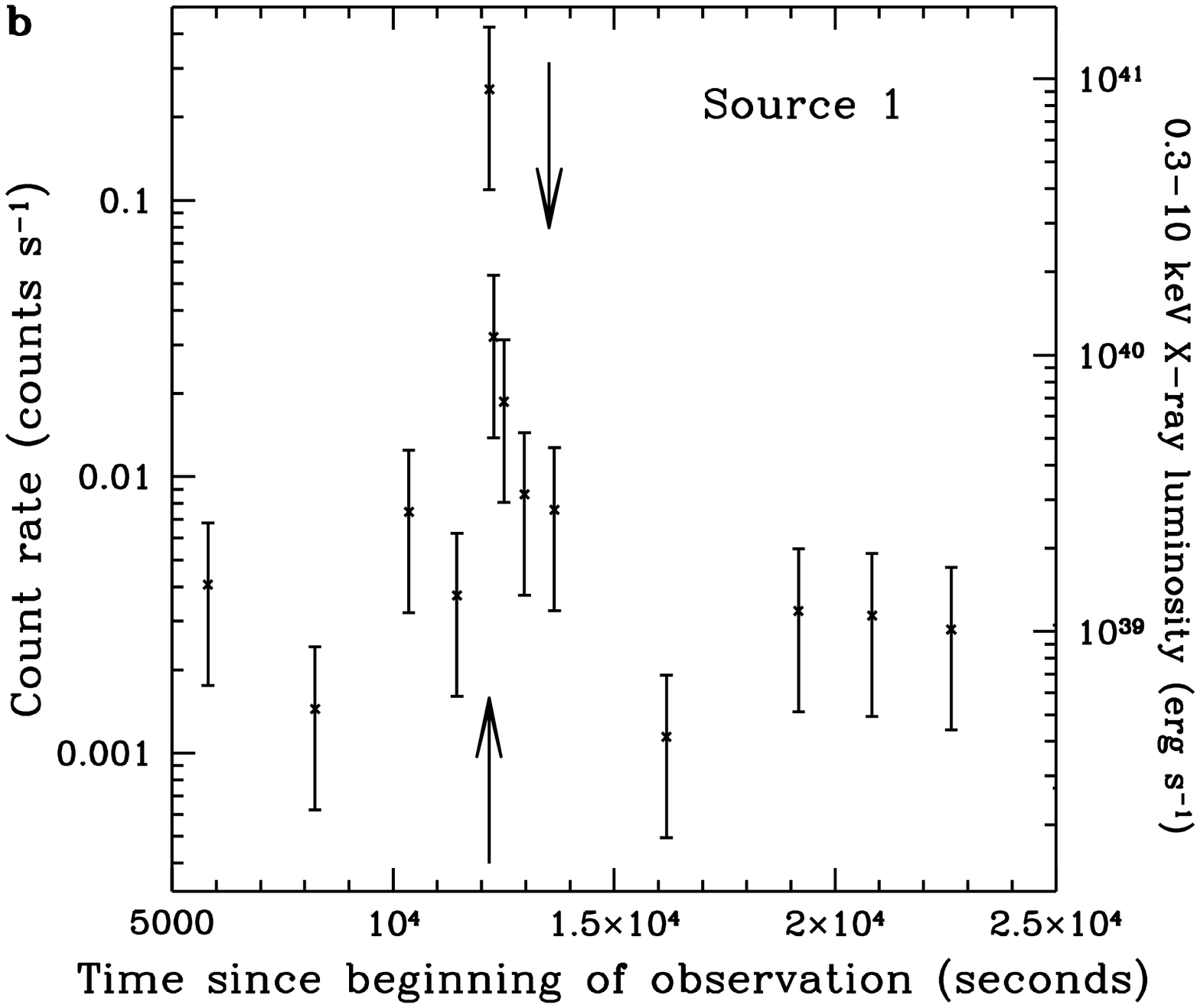}
\end{center}
\caption{\textbf{The {\bfseries{\itshape{Chandra}}} cumulative X-ray photon
arrival time plot and light curve for Source 1 in the NGC~4636 globular
cluster.} {\bf a.}  In total, 162 photons were detected over the $\sim$76,000
second observation. The flare began at the 12,000 second mark of the observation
and lasted for 1,400 seconds. The beginning and ending of the flare are
indicated by up and down arrows, respectively. {\bf b.} Within the gray shaded
region of the cumulative X-ray photon arrival time plot we derive the
background-subtracted X-ray light curve. Each time bin contains five photons,
with error bars representing the 1$\sigma$ uncertainty expected from Poisson
statistics.
\label{fig:ngc4636}}
\end{figure*}

\begin{figure*}
\begin{center}
\includegraphics[width=3.2in]{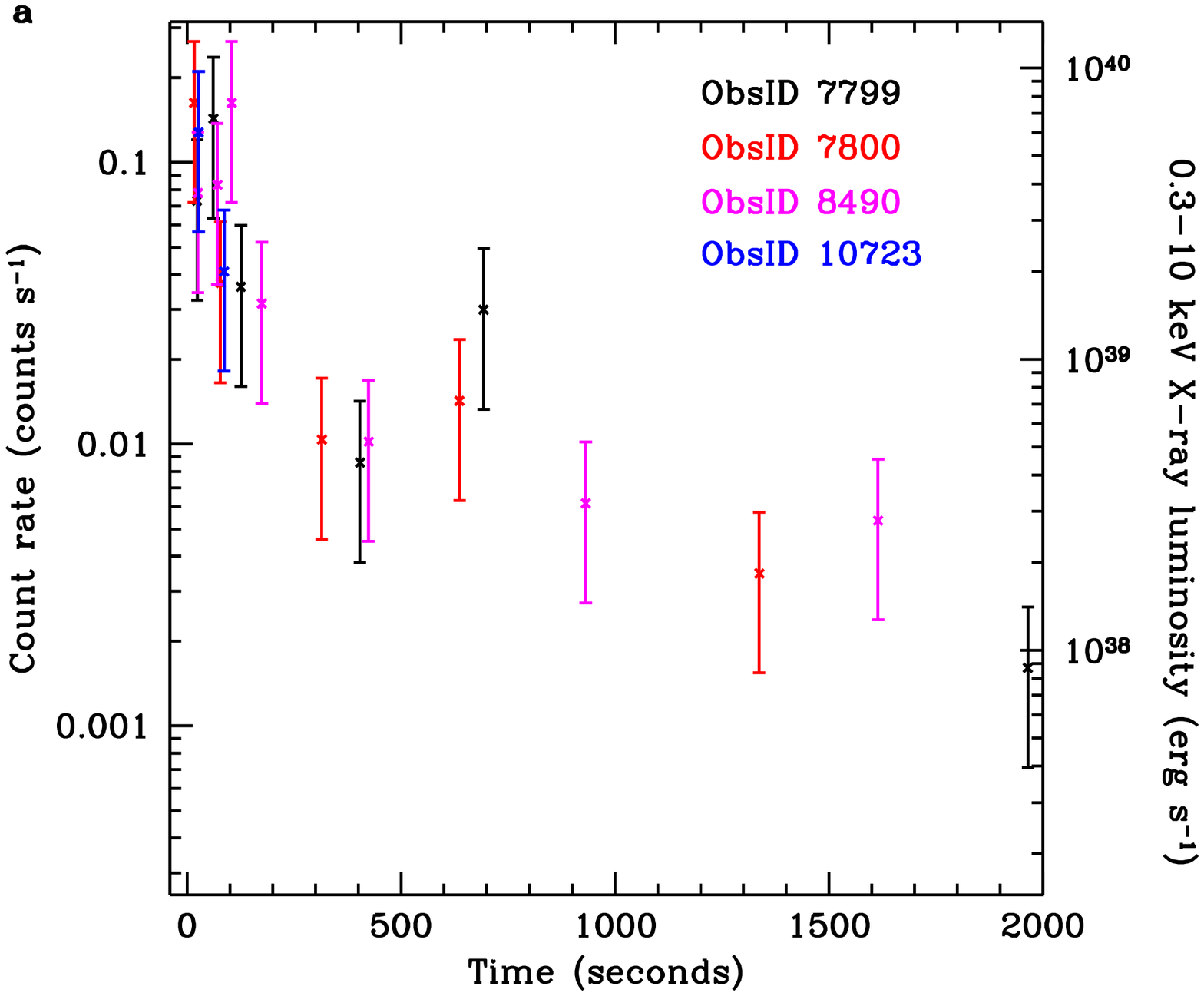}
\includegraphics[width=3.2in]{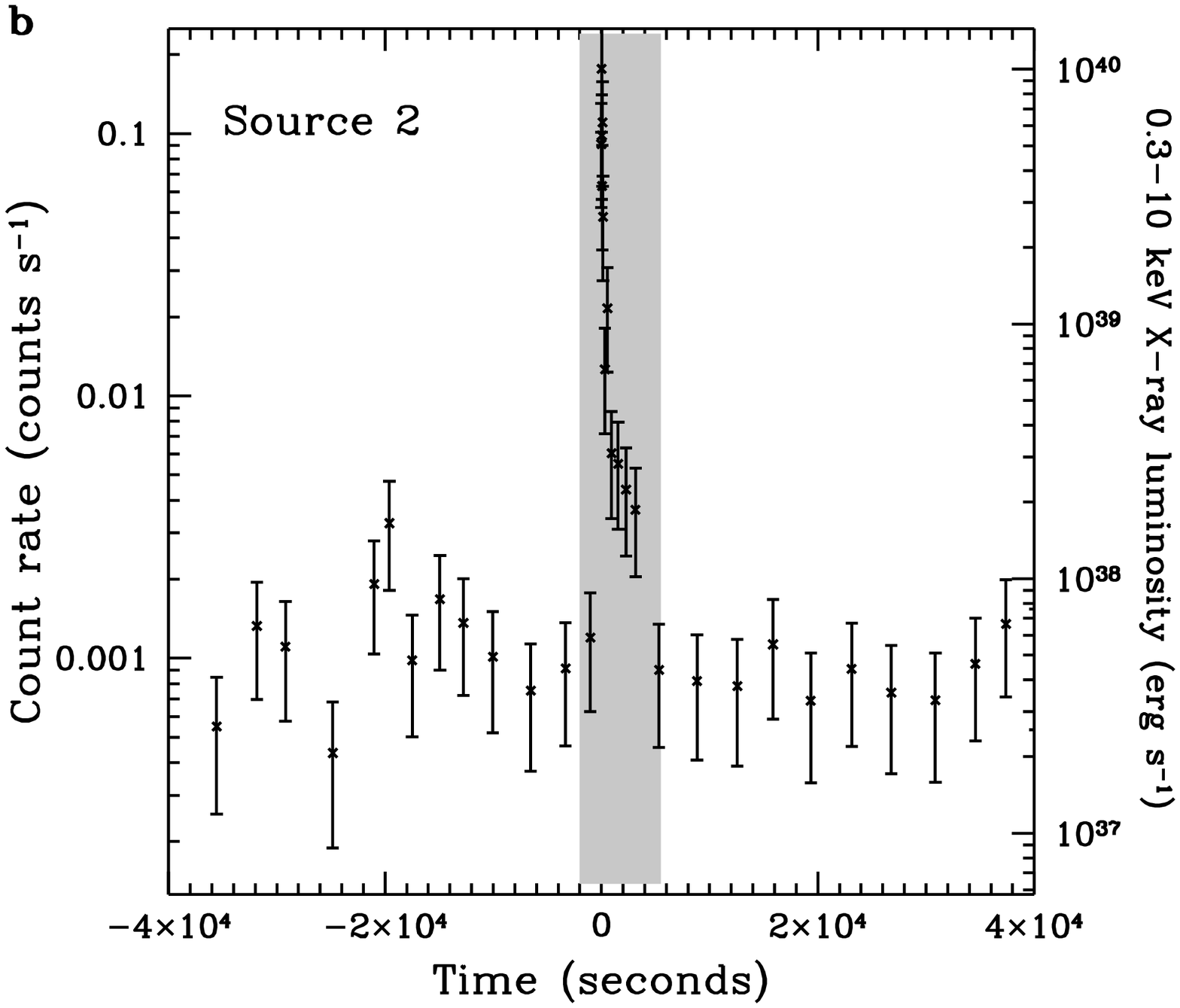}
\includegraphics[width=3.2in]{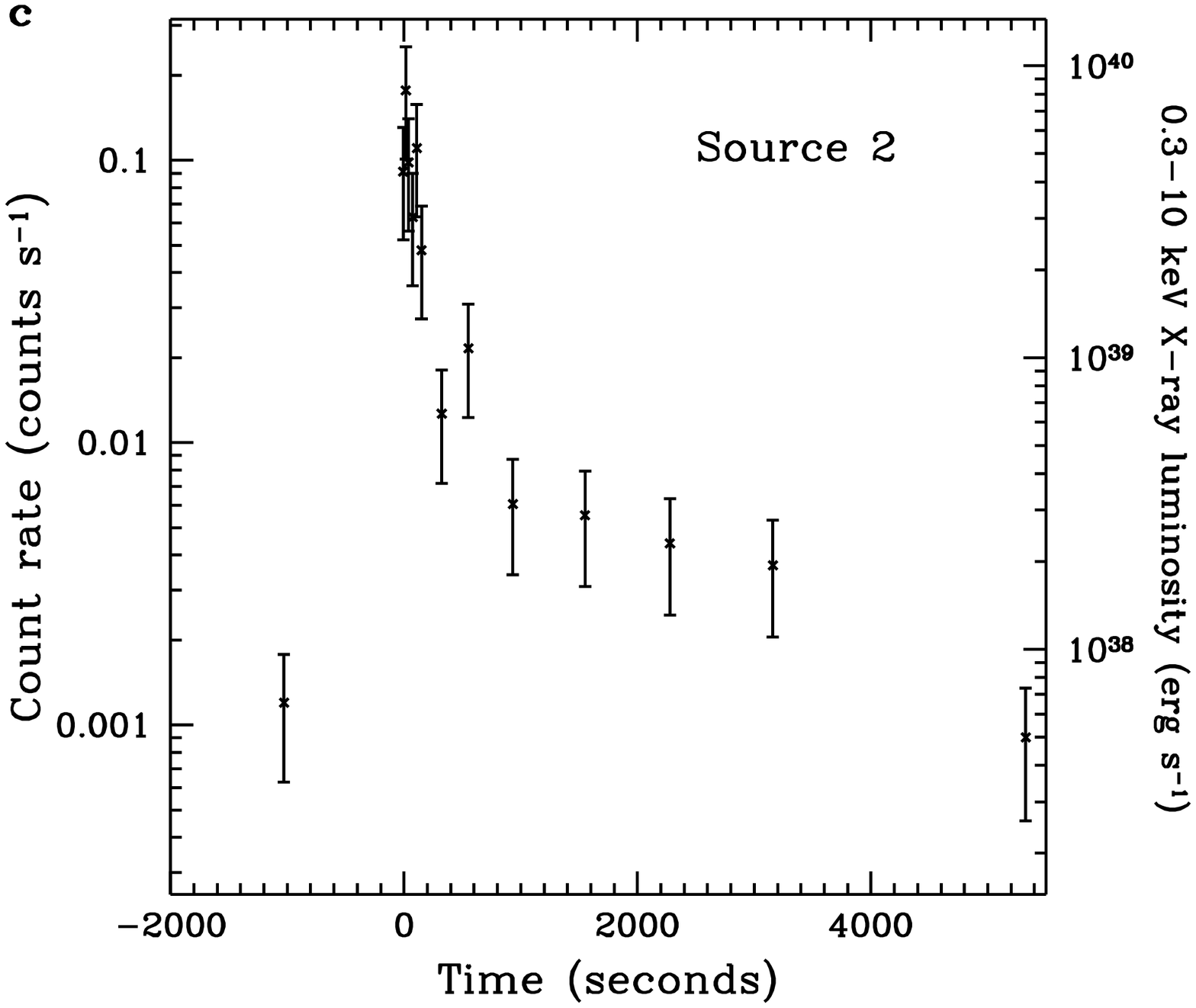}
\end{center}
\caption{\textbf{Individual and combined background-subtracted X-ray light
curves of Source 2 in the NGC~5128 globular cluster/ultracompact dwarf.}
{\bf a.} The X-ray light curves for the four {\it Chandra} flares show similar
behavior. Each time bin contains five photons, with 1$\sigma$ uncertainties.
{\bf b.} The combined light curve of the four flares illustrates the fast
rise/slow decay of the flares. Each time bin contains ten photons. {\bf c.}
Zooming in on the gray shaded region reveals that the luminosity during the
flare rose quickly and remained steady in an ultraluminous state for
$\sim$200 seconds before decaying back to its persistent level after
$\sim$1 hour.
\label{fig:ngc5128}}
\end{figure*}

\clearpage

\section*{Methods}

~
\smallskip

\noindent \textbf{Flare Search Technique}

We searched for flares from all point sources found in 70 {\it Chandra}
observations of nearby luminous early-type galaxies. The {\it evt2} files were
downloaded from the {\it Chandra} archive, and the source detection routine
{\it wavdetect} in the Chandra Interactive Analysis of Observations (CIAO)
package suite on the image files to create a list of sources detected at the
$>$3$\sigma$ level. Our script then extracted the time-ordered photon arrival
times for each source found by {\it wavdetect}. Next, our routines scanned the
photon event list and searched for bursts by finding the time difference
between each photon and the photon three photons forward in time from it (i.e.,
a 4-photon burst) and then calculated the Poisson probability of detecting that
many photons over that time interval given the overall count rate of the source
over the entire observation and the number of 4-photon burst trials present
over the epoch. This was repeated for 5-photon bursts, 6-photon bursts,
7-photon bursts, etc., up to 20-photon bursts. If the probability of a burst of
that magnitude from Poisson fluctuations was below our fiducial value
(1 in $10^6$) and the count rate during the N-photon burst was more than ten
times the average count rate of the source over the observation the source was
marked for further study. Note that our technique is more sophisticated than a
simple KS-test on this distribution. Our technique found several flares from
Milky Way M dwarf stars, both previously known and unknown, which were removed
from consideration. Among the 7,745 sources detected in the 70 observations,
Source 1 and Source 2 were the only non-Galactic sources not previously
detected by other groups (which appear to be transients or one-time
events$^{20-22}$) that were identified for which the random Poisson fluctuation
probability was less than $10^{-6}$, and for which peak--to--persistent count
ratio exceeded ten (to exclude somewhat variable, but non-flaring sources).

\noindent \textbf{{\bfseries{\itshape Chandra} {\rm \bf and}
{\itshape XMM-Newton}} Data Reduction} 

Those {\it Chandra} observations containing flares were then analyzed further
using {\tt CIAO}~4.7 with CALDB version 4.6.9. The sources exhibited flaring
only in ObsIDs 3926 (for Source 1) and in ObsIDs 7799, 7800, 8490, and 10723
(for Source 2). The remaining three and 36 ObsIDs for Source 1 and Source 2,
respectively, showed no flaring activity, did not have the source in the
field of view of the detector, or the data were taken with a lower
sensitivity detector ({\it Chandra} HRC/LETG/HETG) for which a flare of
comparable intensity would not have been detected. Extended Data Table 1
lists all the searched {\it Chandra} and {\it XMM-Newton} observations of
NGC~4636  and NGC~5128. The luminosity of the sources in the non-flare
observations were consistent with the persistent luminosities of the sources
during the flare observations. All flares occurred in ACIS-I pointings. The
event lists were reprocessed using the latest calibration files at the time
of analysis with the {\tt CIAO} tool {\it chandra\_repro}. None of the
{\it Chandra} observations had any background flaring time intervals
significant enough to warrant their removal considering we are interested in
point sources. Energy channels below 0.3 keV and above 6.0 keV were ignored.
Both Sources 1 and 2 were located at least $4\farcm8$ from the ACIS-I aim
point in all the observations, so we used the {\tt CIAO} tool
{\it psfsize\_srcs} to determine the extraction radius for each observation
that enclosed 90\% of the source photons at an energy of 2.3 keV. All
subsequent count rates and 0.3--10 keV X-ray luminosities were corrected for
these point spread function (PSF) losses. Since each flare occurred at a large
off-axis angle from the aim point, the photons were spread over a large PSF.
Thus pile-up effects were negligible even at the peak of each flare. We do not
believe there is any way for the flares to be an instrumental effect such as
pixel flaring or cosmic ray afterglows, for which each recorded event during
the flare would occur in a single detector pixel. Inspection of all the flare
observations in detector coordinates revealed that the photons were not
concentrated in a single detector pixel, but were instead spread out in
detector space in accordance with the dither pattern of {\it Chandra} as
would be expected from astrophysical photons. Furthermore, the photon energies
of cosmic ray afterglows decrease with each successive photon which is not the
case for the photons occurring during the flare.

While we did not conduct a survey of galaxies observed with {\it XMM-Newton}, we
did utilize archival {\it XMM-Newton} data to search for additional flares from
Source 1 and 2. The {\it XMM-Newton} observations for Source 1 did not reveal any
detectable flaring behavior, but the 2014 February 9 (ObsID 0724060801) of
Source 2 revealed a fifth flare for this source detected in both MOS1 and MOS2
detectors separately. To analyze the data, we used the 2014 November 4 release of
the {\it XMM-Newton} Science Analysis Software (SAS), and the data were processed
with the tool {\it emproc}, which filtered the data for the standard MOS event
grades. Source 2 was only observed with the two MOS instruments, since it was not
in the PN camera field of view (the observations used a restricted window due to
the high count rate of the central AGN in NGC 5128). Periods of high background
at the end of the observation were removed.

\noindent \textbf{Cumulative Photon Arrival Time Plots for Sources 1 and 2}

Each X-ray photon collected by {\it Chandra} or {\it XMM-Newton} is tagged with
a position, energy, and time of arrival, allowing a photon-by-photon account of
each X-ray source at a time resolution set by the read-out time for the detector
(3.1 seconds for {\it Chandra} ACIS-I and 2.6 seconds for {\it XMM-Newton} EPIC
MOS). Cumulative photon arrival time plots are a simple way to observe time
variability over the course of the observation. While a source with constant flux
will yield a cumulative photon arrival time plot with a constant slope, a flare
will appear as a nearly vertical rise in the plot as photons stream in over a short
period of time. Figure~1 shows the cumulative photon arrival time plot for Source 1,
illustrating the onset of the flare around the 12,000 second mark following the
beginning of the observation.

The cumulative photon arrival time plot from each of the five flares from Source 2
is shown in Extended Data Figure~1. In each plot the beginning of the flare is
evident. The final {\it Chandra} flare (ObsID 10723) occurred just at the end of
this short observation. The persistent count rate within the 10$^{\prime\prime}$
source extraction region of the {\it XMM-Newton} observation is compromised by
background, but the onset of the flare at the 16,000 second mark is evident.

\smallskip
~

\noindent \textbf{Peak Flare Rate and the Statistical Significance of the Flares}

We estimated the peak flare rate of Source 1 based on the arrival times of the first
six photons of the flare, which arrived over a 22 second period. Given the
uncertainty in when the peak ended, we neglect the sixth photon of the flare to
conservatively estimate a count rate of $0.25^{+0.17}_{-0.11}$ counts s$^{-1}$
(1$\sigma$ uncertainty) after correcting for the 10\% of emission expected to be
scattered out of the source extraction region due to PSF losses. Background was
negligible during the flare, and accounted for only 7\% of the emission inside the
source extraction region during persistent times.

Since Source 1 only flared once, it is necessary to determine accurately how many
independent trials were contained in the sources searched within our sample of
galaxies to determine the likelihood that the flare could result from a random
fluctuation in the persistent count rate. The two sources discussed here were found
as part of a 70 observation sample observed with {\it Chandra} composed primarily of
large elliptical galaxies at distances of 20 Mpc and closer, with a majority of the
galaxies residing in Virgo or Fornax. Within these 70 observations, 7,745 sources were
detected yielding a total of $8.5 \times 10^5$ photons. This is equivalent to
$1.7 \times 10^{5}$ independent 5-photon groupings. Statistically, the chance of
detecting five or more photons in 22 seconds for a source that normally emits at
$1.9 \times 10^{-3}$ counts s$^{-1}$ (the count rate in the persistent state before
correcting for PSF losses) is $1.0 \times 10^{-9}$. With $1.7 \times 10^{5}$
independent trials throughout our sample, the chance of finding a single 5-photon burst
for the Source 1 flare is $1.7 \times 10^{-4}$. Searching over multiple photon burst
scales increases the odds of finding a chance statistical fluctuation. A previous
study$^{1}$ performing a similar calculation has calculated this correction
factor to be $\sim$2.5 using Monte Carlo simulations, so we apply that correction here
leading to a false detection rate of $4.3 \times 10^{-4}$. A similar exercise for the
entire burst (25 photons in 1,400 seconds) leads to a chance fluctuation probability of
$6.4 \times 10^{-6}$.

A similar exercise can be performed for each flare detected in Source 2 by {\it Chandra}
observations. In each case, the flare at its peak was detected using nine photons in
51 seconds, six photons in 22 seconds, seven photons in 22 seconds, and six photons in
37 seconds, respectively. Given the persistent count rates in each observation, and
correcting appropriately for both the number of independent 9-photon, 6-photon, and
7-photon trials (i.e., scaling appropriately from the $1.7 \times 10^5$ independent
5-photon trials) and the multi-burst search correction factor of 2.5, we calculate
probabilities of $1.4\times10^{-5}$, $7.1\times10^{-7}$, $1.2\times10^{-6}$, and
$9.0\times10^{-4}$ of a false flare detection for each flare, respectively. Since we only
considered {\it XMM-Newton} data after having detected the flares in the {\it Chandra}
data, the probability that the flare observed with {\it XMM-Newton} was falsely detected
is $5.0 \times 10^{-8}$ given the 113,000 seconds of total exposure on this source. When
combined this gives a probability that all the flares were falsely detected is
$5.4 \times 10^{-28}$.

\noindent \textbf{X-ray Light Curves}

Owing to the limited photon statistics for the flare in Source 1, only a crude X-ray light
curve was obtained by binning photons in groups of five and determining the count rate over
which the five photons were collected (Figure~1). The four individual 5-photon bin
{\it Chandra} light curves for Source 2 showed similar timing behavior (Figure~2), which gave
us confidence to combine them into one light curve. For each flare, we determined the average
arrival time of the first three photons of the flare and set this to `time zero'. Thus,
photons before the flare were assigned a negative time value. The four photon lists were then
combined at `time zero' to provide a combined photon list. Photons were then binned in groups
of ten to calculate count rates during the time period over which the ten photons were
collected. The count rates were divided by four to give the average count rate per time bin
per flare. Since the fourth {\it Chandra} epoch (ObsID 10723) was very short and does not
extend from --40,000 seconds to 40,000 seconds from the start of the flare, we corrected the
count rate accordingly to account for the temporal coverage of this epoch. The count rates
were corrected for the loss of photons outside the extraction region due to the PSF, and for
the expected background (while negligible during the flare, this accounted for 14\% of the
emission during persistent periods). The combined light curve for Source 2 is shown in
Figure~2. A sharp rise at the beginning of the flare was followed by a flat ultraluminous
state for $\sim$200 seconds. The improvement in statistics by combining the four light curves
traces the duration of the decay in flux out to $\sim$4,000 seconds. Following the flare, the
count rate of the source was remarkably consistent with the pre-flare count rate.

\noindent \textbf{Spectral Fitting and Source Luminosities}

For Source 1, we extracted a combined spectrum during the pre- and post-flare period using
the {\tt CIAO} tool {\tt specextract}. Background was collected from a source-free region
surrounding our source. Using XSPECv12.8, a power law model absorbed by the Galactic column
density in the direction of NGC~4636 ($N_H = 1.8 \times 10^{20}$ cm$^{-2}$)$^{25}$
using the {\sc tbabs} absorption model was employed to fit the background-subtracted spectrum.
Only energy channels over the range 0.5--6.0 keV were considered in the fit. The spectrum was
grouped to contain at least one count per channel and the C-statistic was used in the fit. A
best-fit power law photon index of $1.6\pm0.3$ (90\% uncertainty) was found. This fit implies
an unabsorbed luminosity of $7.9\pm0.8 \times 10^{38}$ erg s$^{-1}$ during the persistent state
(all reported luminosities below have also been corrected for absorption). Since the flare
period contained only 25 photons, the flare spectrum was poorly well-constrained
($\Gamma = 1.6 \pm 0.7$). This led to a peak luminosity during the first 22 seconds of the
flare of $9^{+6}_{-4} \times 10^{40}$ erg s$^{-1}$, a factor of $\sim$120 greater than during
persistent periods combined. Freeing the absorption did not substantially change the fit.
Fitting the flare with a disk blackbody model gave a slightly worse fit with
$kT_{\rm blackbody} = 1.3^{+2.0}_{-0.5}$ keV and a luminosity 30\% less than the power law fit.

For Source 2, we combined the spectra from the flare periods of the four {\it Chandra}
observations using {\tt specextract} into one spectrum. The same was done for the pre- and
post-flare periods combined. The best-fit power law photon indices for persistent and flare
periods assuming a Galactic column density in the direction of NGC~5128
($8.6 \times 10^{20}$ cm$^{-2}$)$^{25}$ were $1.0\pm0.2$ and $1.2\pm0.3$ (90\% uncertainty),
respectively.  Again, this indicates no significant change in the spectral shape during the
flare. These spectral models implied persistent and peak flare luminosities of
$4.4\pm0.3 \times 10^{37}$ erg s$^{-1}$ and $8.1^{+3.5}_{-2.5} \times 10^{39}$ erg s$^{-1}$,
respectively, an increase of $\sim$200 in less than a minute. When we split the flare period
into the flat ultraluminous (first 200 seconds) and decay times (200--4,000 seconds), we
also found no significant spectral evolution. We allowed the Galactic column density $N_H$ to
vary in the fits and found a somewhat softer photon index ($\Gamma=1.6\pm0.6$ in the persistent
state and $\Gamma=1.3\pm0.7$ during the flare) with $N_H = 6^{+5}_{-4} \times 10^{21}$ cm$^{-2}$
for the persistent state and unconstrained below $5 \times 10^{21}$ cm$^{-2}$ during the flare
(90\% uncertainties for two interesting parameters). In both instances, freeing the absorption
only changed the unabsorbed $L_X$ by $<$10\%. The source does not reside in the dust lane of
NGC~5128, so this excess absorption, if real, might be intrinsic to the source. We also fitted
the flare spectrum with a disk blackbody model with fixed $N_H$ at the Galactic value and found
a best-fit temperature of $2.2^{+1.7}_{-0.6}$ keV, with a comparable goodness-of-fit to that of
the power law model and a luminosity 20\% below that derived from the power law fit.
 
For the {\it XMM-Newton} observation, the spectrum and response files were generated using the
standard SAS tasks {\it evselect}, {\it backscale}, {\it arfgen}, and {\it rmfgen}. Since the
count rate during the pre- and post-flare time period is dominated by background (owing to the
much larger extraction region compared to {\it Chandra} and higher background rates), we did not
extract a spectrum for the persistent period. We extracted the background-subtracted flare
spectrum in a 30$^{\prime\prime}$ region around the source and fitted it with the absorbed power
law described above for {\it Chandra} observations. The slope of the power law was
poorly-constrained ($\Gamma=1.5 \pm 0.5$) owing to the low number of photons detected in the
flare, but the slope was consistent with the fit from the co-added {\it Chandra} spectrum. The
peak luminosity of the flare was $1.6^{+1.1}_{-0.7} \times 10^{40}$ erg s$^{-1}$, again consistent
with the {\it Chandra} flares.

\noindent \textbf{Probability of the Flare and Persistent Emission Being From Two Different
Sources}

We have assumed that the persistent and flare emission emanate from a single source within the
globular cluster hosts of Sources 1 and 2, but it is possible that two separate sources in the
same cluster are responsible for the emission. The probability that a globular cluster hosts an
X-ray binary of a particular X-ray luminosity depends on the luminosity of the source$^{26}$,
as well as the properties of the globular cluster such as its mass, concentration, and metal
abundance$^{27, 28}$. From previous work$^{27}$, the number of X-ray sources more luminous
than $3.2 \times 10^{38}$ erg s$^{-1}$ in a globular cluster that has a mass $M$, stellar
encounter rate $\Gamma_h$, half-light radius $r_h$, and cluster metallicity $Z$ is
$0.041\left( \frac{\Gamma_h}{10^7} \right)^{0.82 \pm 0.05} \, \left(\frac{Z}{Z_\odot}\right)^{0.39\pm 0.07}$,
where $\Gamma_h = \left( (\frac{M}{M_\odot} ) \frac{1}{2\pi} \right)^{3/2} \left( \frac{r_{h}}{1 {\rm \, pc}} \right)^{-5/2}$.

The globular cluster hosting Source 1 has photometry in Kron-Cousins $R$-band and Washington $C$
filters; $R= 23.02$ and  $C-R = 1.94$. This color corresponds to a photometrically derived
metallicity of $Z$/H = --0.08 dex ($Z$ = 0.8 $Z_\odot$)$^{29}$. Using a single population
model$^{30}$ given a Kroupa initial mass function, 13 Gyr age, and $Z$/H = --0.08 dex, an $M/L$
of 4.1 in $R$-band is expected for this cluster. Given the distance to NGC 4636
($d=14.3{\rm \, Mpc}$), the $R$-band $M/L$ referenced above, and $M_{R,\odot}=4.42$, we estimate
a globular cluster mass of $3.0\times 10^{5}$ M$_\odot$. Since we have no size measurement for
this globular cluster, we conservatively estimate a minimum size of 1.5 pc, which is the
3$\sigma$ lower limit based on a survey of globular clusters in the Virgo cluster$^{31}$. With
these values, we estimate the globular cluster is expected to have 0.017 X-ray binaries above
$3.2\times10^{38}$ erg s$^{-1}$. To determine the number of X-ray binaries expected above the
observed persistent X-ray luminosity of Source 1, we apply the X-ray luminosity function in
globular clusters found in a previous study$^{26}$, which predicts that the Source 1 persistent
luminosity ($8 \times 10^{38}$ erg s$^{-1}$) is ten times less likely to be found in a globular
cluster than a $3.2\times10^{38}$ erg s$^{-1}$ source. This leads to an estimate of 0.0017 X-ray
sources equal to or more luminous than Source 1. Thus, after having found a flaring source, the
probability that the persistent emission comes from a different X-ray binary in this cluster is
$<$0.17\%. If we conservatively assume that the predicted number of LMXBs could be 50\% higher
(approximately convolving all of the uncertainty sources), the probability is $<$0.24\%.

The globular cluster/ultracompact dwarf galaxy hosting Source 2 has a spectroscopically-determined
metallicity of $Z$/H = --0.85 dex ($Z$ = 0.14 $Z_\odot$)$^{32}$. The derived stellar mass$^{12}$ of
the source is $3.1\times 10^6$ M$_{\odot}$. Given its size$^{12}$ of 7 pc, and correcting for the
luminosity function$^{26}$ (which predicts that a $4 \times 10^{37}$ erg s$^{-1}$ source is ten
times more likely to be found in a globular cluster than a $3.2 \times 10^{38}$ erg s$^{-1}$
source), we estimate the globular cluster is expected to have 0.064 X-ray binaries above
$4 \times10^{37}$ erg s$^{-1}$. Thus after having found a flaring source, the probability that the
persistent emission comes from a different X-ray binary in this cluster is $<$6.4\%. If we
conservatively assume that the predicted number of LMXBs could be 50\% higher (approximately
convolving all of the uncertainty sources), the probability is $<$9.1\%. We note that this might be
an overestimate given that ultracompact dwarfs appear to harbor X-ray sources at a lower rate than
globular clusters$^{33}$.

Even in the most conservative case, the combined probability that both sources arise from different
sources than the persistently emitting sources is $<1.5\times10^{-4}$.

For both sources, we determined the position of the sources separately during its flare phase and
persistent phase and found no statistical difference within the positional uncertainties. This is
not highly constraining, however, given the large PSF of {\it Chandra} at the off-axis location of
the flares.

\noindent \textbf{Code availability:} 
The code to find X-ray flares is available at  \\
 http://pages.astronomy.ua.edu/jairwin/software/.

\noindent \textbf{References}

\noindent [25.] Dickey, J. M. \& Lockman, F. J. H I in the Galaxy. Annu. Rev. Astron.
Astrophys. 28, 215--261 (1990).

\noindent [26.] Zhang, Z. et al. Luminosity functions of LMXBs in different stellar
environments. Mon. Not. R. Astron. Soc. 533, A33 (2011).

\noindent [27.] Sivakoff, G. R. et al. The Low-Mass X-Ray Binary and Globular Cluster
Connection in Virgo Cluster Early-Type Galaxies: Optical Properties. Astrophys. J. 660,
1246--1263 (2007).

\noindent [28.] Kundu, A., Maccarone, T. J. \& Zepf, S. E. Probing the Formation of
Low-Mass X-Ray Binaries in Globular Clusters and the Field. Astrophys. J. 662,
525--543 (2007).

\noindent [29.] Harris, W. E. \& Harris, G. L. H. The Halo Stars in NGC 5128. III.
An Inner Halo Field and the Metallicity Distribution. Astron. J. 123,
3108--3123 (2002).

\noindent [30.] Maraston, C. Evolutionary population synthesis: models, analysis of
the ingredients and application to high-z galaxies. Mon. Not. R. Astron. Soc.
362, 799--825 (2005).

\noindent [31.] J\'ordan, A. et al. The ACS Virgo Cluster Survey. X. Half-Light
Radii of Globular Clusters in Early-Type Galaxies: Environmental Dependencies
and a Standard Ruler for Distance Estimation. Astrophys. J. 634, 1002--1019 (2005).

\noindent [32.] Beasley, M. A. et al. A 2dF spectroscopic study of globular clusters
in NGC 5128: probing the formation history of the nearest giant elliptical.
Mon. Not. R. Astron. Soc. 386, 1443--1463 (2008).

\noindent [33.] Pandya, V., Mulchaey, J. \& Greene, J. E. A Comprehensive Archival
Chandra Search for X-Ray Emission from Ultracompact Dwarf Galaxies.
Astrophys. J. 819, 162 (2016).

\clearpage

\begin{table*}
\begin{center}
\begin{tabular}{*{10}{c}}
\hline
\hline
Source  & Telescope/Detector & ObsID & Observation Date & Exposure (ksec) & Flare? \\
\hline
1 & {\it Chandra}/ACIS-I & 324 & 1999-12-04 & 8.5 & N \\
1 & {\it Chandra}/ACIS-S & 323 & 2000-01-26 & 53.1 & N \\
1 & {\it XMM-Newton}/EPIC & 0111190101 & 2000-07-13 & 27.2 & N \\
1 & {\it XMM-Newton}/EPIC & 0111190501  & 2000-07-13 & 6.6 & N \\
1 & {\it XMM-Newton}/EPIC & 0111190201 & 2000-07-13 & 66.3 & N \\
1 & {\it XMM-Newton}/EPIC & 0111190701 & 2001-01-05 & 64.4 & N \\
1 & {\it Chandra}/ACIS-I & 3926 & 2003-02-15 & 75.7 & Y \\
1 & {\it Chandra}/ACIS-I & 4415 & 2003-02-15 & 75.3 & N \\
2 & {\it Chandra}/HRC-I & 463 & 1999-09-10 & 19.7 & N \\
2 & {\it Chandra}/HRC-I & 1253 & 1999-09-10 & 6.9 & N \\
2 & {\it Chandra}/ACIS-I & 316 & 1999-12-05 & 36.2 & $\cdots$ \\
2 & {\it Chandra}/HRC-I & 1412 & 1999-12-21 & 15.1 & N \\
2 & {\it Chandra}/HRC-I & 806 & 2000-01-23 & 65.3 & N \\
2 & {\it Chandra}/ACIS-I & 962 & 2000-05-17 & 37.0 & N \\
2 & {\it XMM-Newton}/EPIC & 0093650201 & 2001-02-02 & 23.9 & N \\
2 & {\it Chandra}/ACIS-S/HETG & 1600 & 2001-05-09 & 47.5 & N \\
2 & {\it Chandra}/ACIS-S/HETG & 1601 & 2001-05-21 & 52.2 & N \\
2 & {\it XMM-Newton}/EPIC & 0093650301 & 2002-02-06 & 15.3 & N \\
2 & {\it Chandra}/ACIS-S & 2978 & 2002-09-03 & 45.2 & $\cdots$ \\
2 & {\it Chandra}/ACIS-S & 3965 & 2003-09-14 & 50.2 & $\cdots$ \\
2 & {\it Chandra}/ACIS-I & 7797 & 2007-03-22 & 98.2 & N \\
2 & {\it Chandra}/ACIS-I & 7798 & 2007-03-27 & 92.0 & N \\
2 & {\it Chandra}/ACIS-I & 7799 & 2007-03-30 & 96.0 & Y \\
2 & {\it Chandra}/ACIS-I & 7800 & 2007-04-17 & 92.1 & Y \\
2 & {\it Chandra}/ACIS-I & 8489 & 2007-05-08 & 95.2 & N \\
2 & {\it Chandra}/ACIS-I & 8490 & 2007-05-30 & 95.7 & Y \\
2 & {\it Chandra}/ACIS-I & 10723 & 2009-01-04 & 5.2 & Y \\
2 & {\it Chandra}/ACIS-I & 10724 & 2009-03-07 & 5.2 & N \\
2 & {\it Chandra}/HRC-I & 10407 & 2009-04-04 & 15.2 & N \\
2 & {\it Chandra}/ACIS-I & 10725 & 2009-04-26 & 5.0 & N \\
2 & {\it Chandra}/ACIS-I & 10726 & 2009-06-21 & 5.2 & N \\
2 & {\it Chandra}/ACIS-S & 10722 & 2009-09-08 & 50.0 & $\cdots$ \\
2 & {\it Chandra}/HRC-I & 10408 & 2009-09-14 & 15.2 & N \\
2 & {\it Chandra}/ACIS-I & 11846 & 2010-04-26 & 4.8 & N \\
2 & {\it Chandra}/ACIS-I & 11847 & 2010-09-16 & 5.1 & $\cdots$ \\
2 & {\it Chandra}/ACIS-I & 12155 & 2010-12-22 & 5.1 & N \\
2 & {\it Chandra}/ACIS-I & 12156 & 2011-06-22 & 5.1 & N \\
2 & {\it Chandra}/ACIS-I & 13303 & 2012-04-14 & 5.6 & N \\
2 & {\it Chandra}/ACIS-I & 13304 & 2012-08-29 & 5.1 & N \\
2 & {\it Chandra}/ACIS-I & 15294 & 2013-04-05 & 5.1 & N \\
2 & {\it XMM-Newton}/EPIC & 0724060501 & 2013-07-12 & 12.0 & N \\
2 & {\it XMM-Newton}/EPIC & 0724060601 & 2013-08-07 & 12.0 & N \\
2 & {\it Chandra}/ACIS-I & 15295 & 2013-08-31 & 5.4 & N \\
2 & {\it XMM-Newton}/EPIC & 0724060701 & 2014-01-06 & 26.5 & N \\
2 & {\it XMM-Newton}/EPIC & 0724060801 & 2014-02-09 & 23.4 & Y \\
2 & {\it Chandra}/ACIS-I & 16276 & 2014-04-24 & 5.1 & N \\
2 & {\it Chandra}/ACIS-I & 16277 & 2014-09-08 & 5.4 & $\cdots$ \\
2 & {\it Chandra}/ACIS-I & 17471 & 2015-03-14 & 5.4 & N \\
2 & {\it Chandra}/ACIS-S/LETG & 17147 & 2015-05-13 & 49.7 & N \\
2 & {\it Chandra}/ACIS-S/LETG & 17657 & 2015-05-17 & 50.4 & N \\
\hline
\end{tabular}
\end{center}
\normalsize
{\textbf{\bf Extended Data Table 1: Summary of the {\bfseries{\itshape{Chandra}}}
and {\bfseries{\itshape{XMM-Newton}}} Observations of Source 1 and 2}}
\end{table*}

\begin{figure*}
\vspace{-0.7truein}
\small
\begin{center}
\includegraphics[width=3.0in,angle=0]{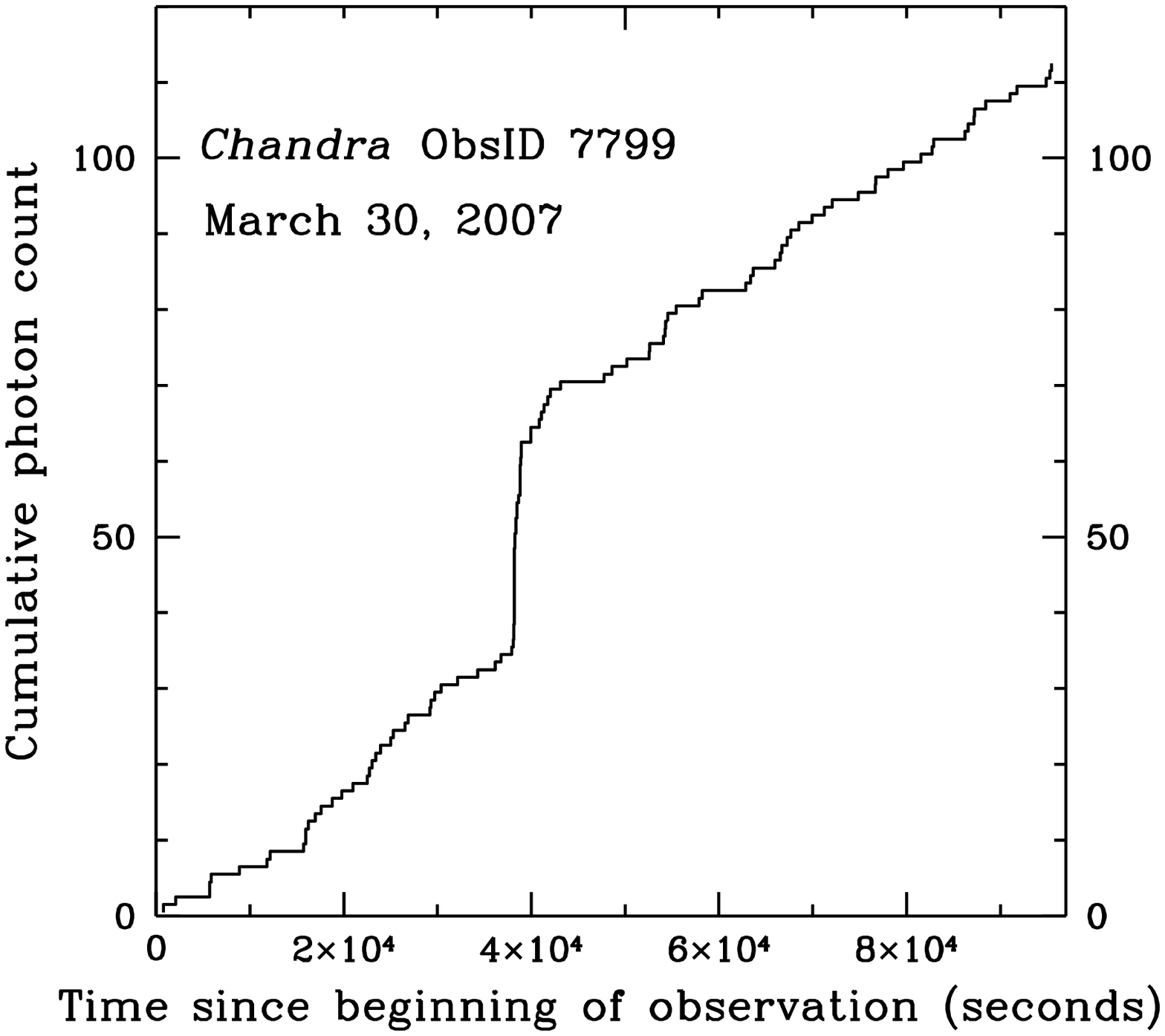}
\includegraphics[width=3.0in,angle=0]{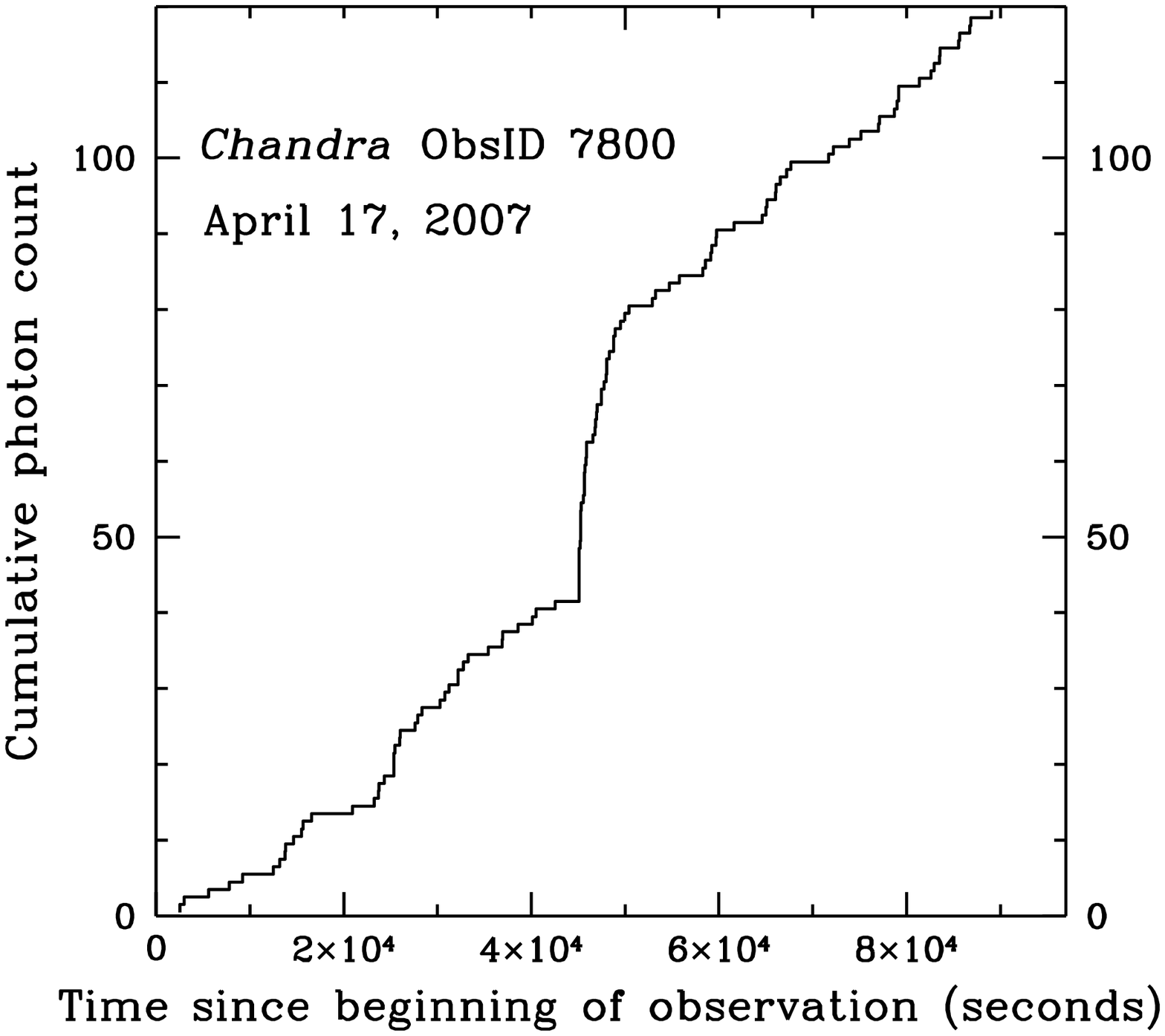}
\includegraphics[width=3.0in,angle=0]{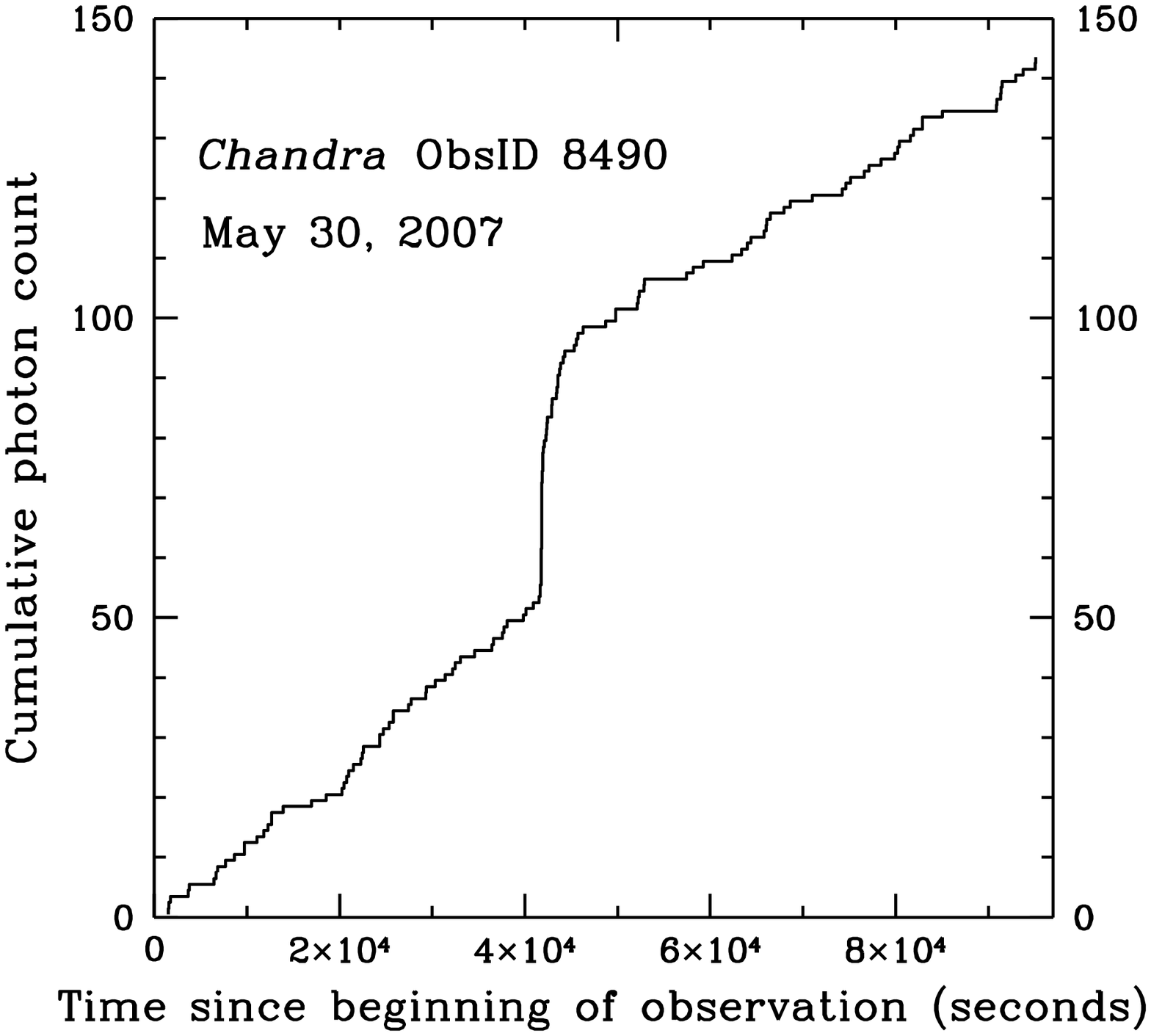}
\includegraphics[width=3.0in,angle=0]{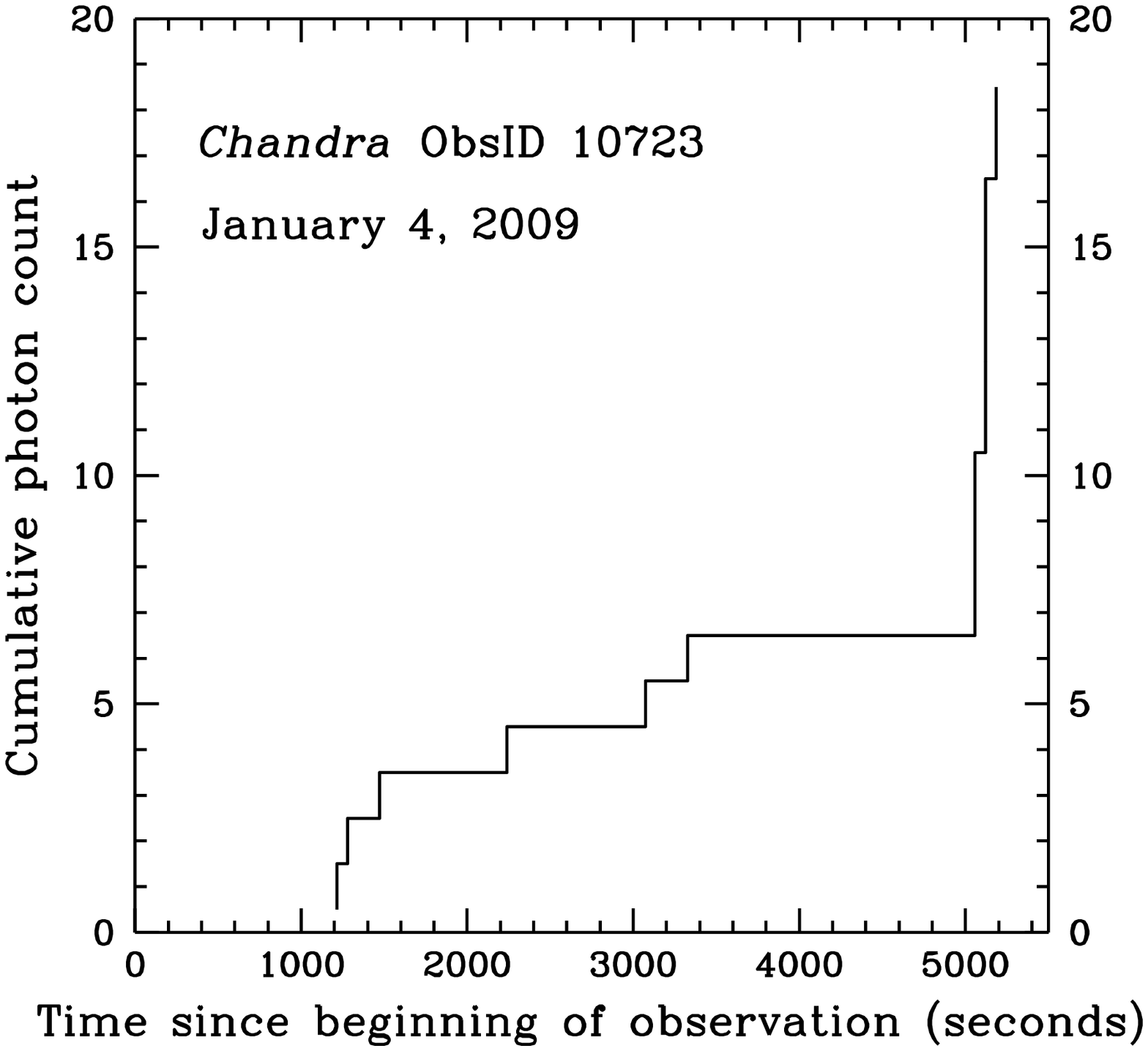}
\includegraphics[width=3.0in,angle=0]{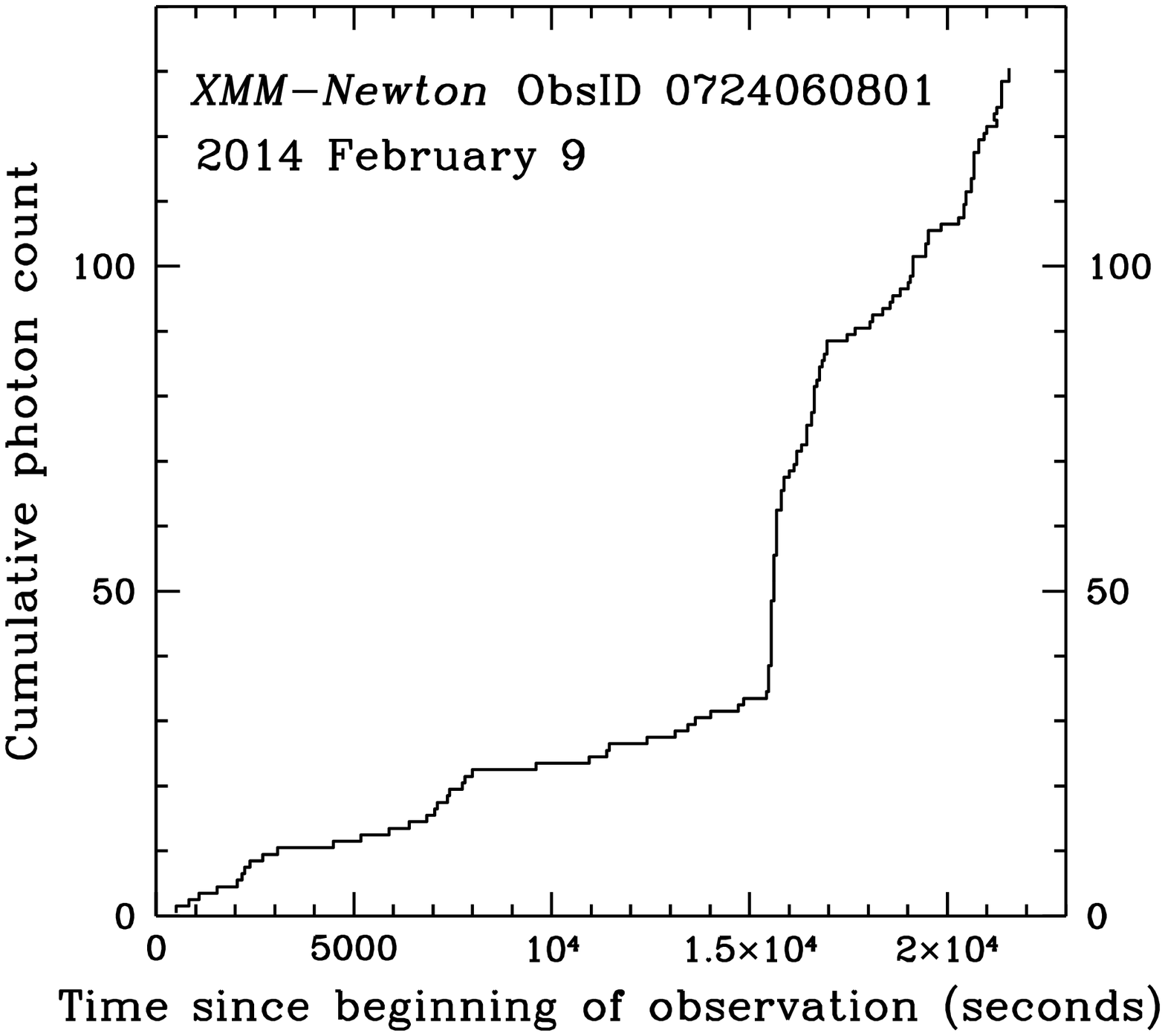}
\end{center}
\vskip -0.2in
{\textbf{Extended Data Figure~1: The cumulative X-ray photon arrival time plots
for the five flares of Source 2 in NGC~5128.} The first four flares were observed by
{\it Chandra} with the fifth flare by {\it XMM-Newton}. In ObsID 10723, the first
photon of the observation was not received until 1100 seconds after
the observation began, and the observation ended mid-flare.
}
\end{figure*}

\end{document}